\documentclass[pre,showpacs,twocolumn,groupedaddress,floatfix,a4paper,10point]{revtex4}
\usepackage[utf8]{inputenc}
\usepackage{epsfig}
\usepackage[T1]{fontenc}
\usepackage{graphicx}
\usepackage{amssymb}
\usepackage{amsmath}

\long\def\beginpgfgraphicnamed#1#2\endpgfgraphicnamed{\includegraphics{#1}}

\usepackage{relsize}
\usepackage{color}
\usepackage{times}
\usepackage{bm}
\usepackage[section]{placeins}
\usepackage{array}
\usepackage[left=1.5cm,top=2cm, bottom=2cm]{geometry}

\usepackage{mathpazo}
\usepackage{helvet}
\usepackage[euler-digits, euler-hat-accent]{eulervm}
\usepackage[mathscr] {eucal}
\usepackage{dsfont}
\newcommand{\mbold}[1]{\mathbold{#1}}

\usepackage{ragged2e}
\usepackage[compatibility=false,font=small]{caption}
\DeclareCaptionJustification{reallyjustified}{\justifying}
\DeclareCaptionStyle{default}{justification=reallyjustified,font=small}
\usepackage[justification=justified]{subfig}

\usepackage{placeins}

\newcommand{\avg}[1]{{\left<#1\right>}}

\begin{document}

\title{Boolean networks with reliable dynamics}

\author{Tiago P. Peixoto}
\email[]{tiago@fkp.tu-darmstadt.de}
\author{Barbara Drossel}
\email[]{drossel@fkp.tu-darmstadt.de}
\affiliation{Institut für Festkörperphysik, TU Darmstadt, Hochschulstrasse 6,
  64289 Darmstadt, Germany}

\date{\today}

\begin{abstract}
  We investigated the properties of Boolean networks that follow a given
  reliable trajectory in state space. A reliable trajectory is defined as a
  sequence of states which is independent of the order in which the nodes are
  updated. We explored numerically the topology, the update functions, and the
  state space structure of these networks, which we constructed using a minimum
  number of links and the simplest update functions. We found that the
  clustering coefficient is larger than in random networks, and that the
  probability distribution of three-node motifs is similar to that found in gene
  regulation networks. Among the update functions, only a subset of all possible
  functions occur, and they can be classified according to their
  probability. More homogeneous functions occur more often, leading to a
  dominance of canalyzing functions. Finally, we studied the entire state space
  of the networks. We observed that with increasing systems size, fixed points
  become more dominant, moving the networks close to the frozen phase.
\end{abstract}


\maketitle

\section{Introduction}
Boolean networks (BNs) are used to model the dynamics of a wide variety of
complex systems, ranging from neural networks \cite{rosen-zvi_multilayer_2001}
and social systems \cite{moreira_efficient_2004} to gene regulation networks
\cite{lagomarsino_logic_2005}.  BNs are composed of nodes with binary states,
coupled among each other.  The state of each node evolves according to a
function of the states from which it receives its inputs, similarly to what is
done when using cellular automata~\cite{wolfram_new_2002}, but in contrast to
cellular automata, BNs have no regular lattice structure, and not all nodes are
assigned the same update function.

The simplest type of BNs are random BNs~\cite{kauffman_metabolic_1969}, where
the connections and the update functions are assigned at random to the
nodes. These random models have the advantage of being accessible to analytical
calculations, thus permitting a deep understanding of such systems
\cite{drossel_random_2008}. Random BNs can display three types of dynamical
behavior, none of which is very realistic: in the ``frozen'' phase, most or all
nodes become fixed in a state which is independent of the initial conditions. In
the ``chaotic'' phase, attractors of the dynamics are extremely long, and
dynamics is very sensitive to perturbations. At the critical point between these
two phases, attractor numbers are huge and depend strongly on the update scheme
used~\cite{greil_dynamics_2005,klemm_stable_2005}.

In contrast to random BNs, real biological networks typically display a highly
robust behavior. For instance, the main dynamical trajectory of the yeast
cell-cycle network model derived by Li et al. \cite{li_yeast_2004} changes
little when the nodes are updated in a different order, and the system returns
quickly to this trajectory after a perturbation.  In fact, whenever the
functioning of a system depends on the correct execution of a given sequence of
steps, the system must be robust with respect to the omnipresent effects of
noise.

Motivated by this requirement, we focus in the present paper on the
robustness of dynamical trajectories under fluctuations in the time at
which the nodes are updated. We consider the extreme case, where we
require the system to have a trajectory that is completely robust under a
change in the update sequence. This means that at any time all but one
node would remain in their present state when they are updated.

In contrast to the standard approach to BNs, where first the network structure
(i.e., the topology and update functions) is defined and then the dynamics is
investigated, we define first the dynamical trajectory and then construct
networks that satisfy this trajectory, with the trajectory being robust under
changes in the update sequence.  A similar method has been used in
\cite{lau_function_2006}. In the next section, we will define the model and
methods used. Then, we will discuss the properties of the networks constructed
by this methods, considering the topology, the update functions, and the state
space structure.  Finally, we will outline directions for further
investigations.

\section{The model}\label{sec:model}

A BN is defined as a directed network of $N$ nodes representing
Boolean variables $\mbold{\sigma} \in \{1,0\}^N$, which are subject to a
dynamical update rule,
\begin{equation}
  \sigma_i(t+1) = f_i\left(\mbold{\sigma}(t)\right) u_i(t) +
  \sigma_i(t)\left[1-u_i(t)\right]
\end{equation}
where $f_i$ is the update function assigned to node $i$, which depends
exclusively on the states of its inputs. The binary vector $\mbold{u}(t)$
represents the \emph{update schedule}, and has components $u_i(t) = 1$ if node
$i$ should update at time $t$, or $u_i(t) = 0$ if it should retain the same
value. The update functions $f_i$ are conveniently indexed by the outputs of
their truth table as follows: Given an arbitrary input ordering, each input
value combination $\mbold{\sigma}_j=\{\sigma_0,\sigma_1,\dots,\sigma_{k-1}\}$
will have an associated index $c(\mbold{\sigma})=\sum_i\sigma_i2^i$ which
uniquely identifies it. Any update function $f$ can in turn be uniquely indexed
by $f=\sum_jf(\mbold{\sigma}_j)2^{c(\mbold{\sigma}_j)}$, where
$f(\mbold{\sigma}_j)$ is the output of the indexed function, given the input
value combination $\mbold{\sigma}_j$.

The update schedule can be chosen in three different ways: (a) Synchronous
(parallel update), where $\mbold{u}(t)=\mathbb{1}$, and all nodes are updated
simultaneously every time step; (b) Asynchronous and deterministic, where, for
instance, $\mbold{u}(t)=\{1 - \Theta((t+t^0_i) \mod t_i)\}$, where $t_i$ is the
period with which vertex $i$ is updated, $t^0_i$ is a local phase, and
$\Theta(x)$ is the Heaviside step function; and finally (c) Asynchronous and
stochastic, where $u_j=1$ and $u_{i\neq j} = 0$; in the fully stochastic case
$j$ is a random value in the range $[1,N]$, chosen independently at each time
step.

The choice of update schedule should take into account the fact that processes
in biological (cellular) networks are subject to stochastic fluctuations which
can affect the timing of the different steps. In principle, a network could be
organized such that the time interval between subsequent updates is so large
that the update sequence is not affected by a small level of noise in the update
times. In this case, an asynchronous deterministic updating scheme would be
appropriate. However, more generally, the noise will also affect the sequence in
which nodes are updated, suggesting an updating scheme that contains some degree
of stochasticity.

\begin{figure}[hbt]
  \subfloat[Stochastic dynamics]{\includegraphics*[width=0.45\columnwidth]{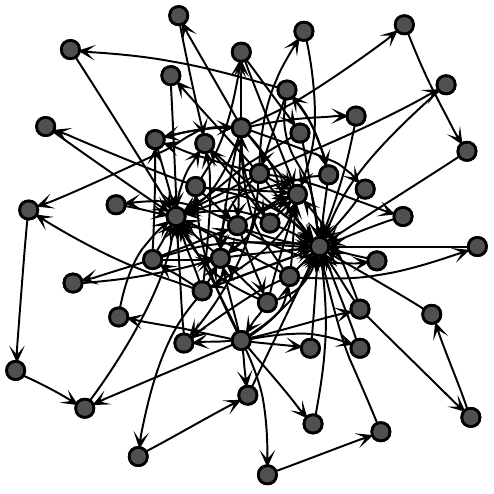}\label{fig:stochastic}}
  \subfloat[``Checkpoint'' states]{\includegraphics[width=0.45\columnwidth]{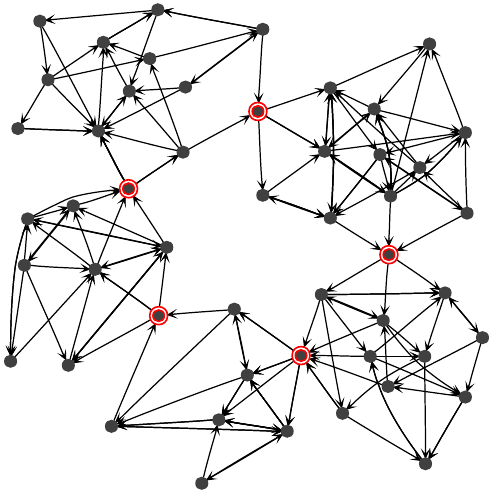}\label{fig:checkpoints}}\\
  \subfloat[Entirely reliable trajectory]{\includegraphics*[width=0.45\columnwidth]{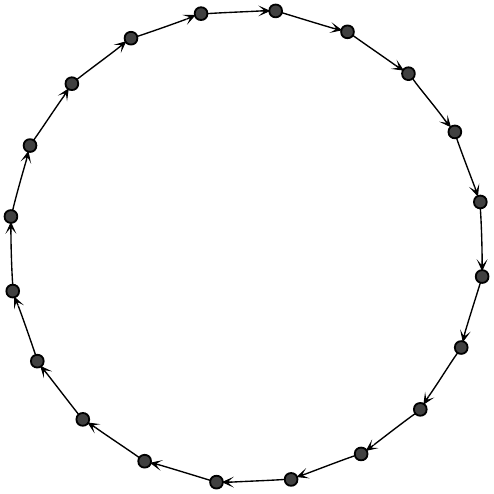}\label{fig:reliable}}
  \caption{(Color online) Ilustration of levels of dynamical reliability. Each
    node on the graphs above is a state of the system, and the edges represent
    possible transitions between them.\label{fig:reliability}}
\end{figure}

In principle, networks can respond in different ways to stochasticity in the
update sequence (see Fig.~\ref{fig:reliability}): (a) The system has no specific
sequence of states, and it quickly looses memory of its past states. (b) The
system has some degree of ordering in the sequence of states, with
``checkpoint'' states that occur in a given order, and with certain groups of
states occurring in between. (c) The system has entirely reliable dynamics,
where the sequence of states is always the same on the attractor, no matter in
which order the nodes are updated.

In this paper, we will focus on systems that have an attractor that has entirely
reliable dynamics. Many cellular processes, such as the response to some
external signal, or the cell cycle, can only function correctly if the system
goes through the required sequence of states in the correct order. Therefore,
considering the idealization of fully reliable dynamics is biologically
motivated.  Furthermore, studying networks with entirely reliable dynamics is
also of theoretical interest, since it is an idealized situation on which one
can build when studying more complicated cases. Entirely reliable dynamics can
be implemented by enforcing that consecutive states of the attractor trajectory
differ only in the value of one node. In other words, the Hamming distance
between successor states is always 1. It is obvious that this is the only
possible type of trajectory that can be entirely independent of the update
schedule. If two subsequent states differed by the state of two or more nodes,
then it would be possible to devise an update sequence which would update one
node but not the other, in contradiction to our assumption.

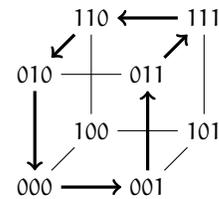
\begin{figure}[htb]
  \beginpgfgraphicnamed{fig-cube}
  \begin{tikzpicture}
    \node (000) at (0,0) [] {$000$};
    \node (001) at (1.5,0) [] {$001$} edge [<-,very thick] (000);
    \node (010) at (0,1.5) [] {$010$} edge [->,very thick] (000);
    \node (011) at (1.5,1.5) [] {$011$} edge [-] (010) edge [<-,very thick] (001);

    \node (100) at (0.75,0.75) [] {$100$} edge [-] (000);
    \node (101) at (2.25,0.75) [] {$101$} edge [-] (100) edge [-] (001);
    \node (110) at (0.75,2.25) [] {$110$} edge [->, very thick] (010) edge [-] (100);
    \node (111) at (2.25,2.25) [] {$111$} edge [->, very thick] (110) edge [<-,very thick] (011) edge [-] (101);
  \end{tikzpicture}
  \endpgfgraphicnamed
  \caption{Example of reliable trajectory of length $6$ on a system of size
    $N=3$.\label{fig:cube}}
\end{figure}

Entirely reliable attractors are represented in state space as simple loops. We
denote the number of different states on the attractor by $L=\sum_il_i$, where
$l_i$ is the number of times node $i$ changed its state during a full period
(since the trajectory is periodic, $l_i$ must be equal to 0 or a multiple of
2). Furthermore, if the states of the system were represented by the corners of
a $N$-dimensional Hamming hypercube, the trajectory should follow its edges (see
Fig.~\ref{fig:cube}). The shortest possible trajectory length, considering that
no node remains at a constant value, is $L=2N$, with $l_i=2$ for all nodes. The
longest possible trajectory length is $L=2^N$, where all states of the system
are visited, and the trajectory corresponds to a Hamiltonian walk on the
$N$-dimensional Hamming hypercube~\footnote{Unlike in some types of graphs, those
  trajectories are always possible on hypercubes, and are known as Grey codes in
  computer science~\cite{knuth_art_2005}.}.

\section{Minimal reliable networks}\label{sec:minimal}
\subsection{Construction rule}
The goal of this section is to construct BNs that have a given entirely reliable
trajectory, and to investigate their properties. A fully reliable trajectory has
the property that the sequence of states is independent of the updating scheme,
which means that under parallel update only one node at a time changes its
state. How networks that go through a given sequence of states can be
constructed, was demonstrated by Lau et al~\cite{lau_function_2006}, who
investigated all possible networks which exhibit the main trajectory of the
Yeast cell cycle regulatory network. Thus, we first define the dynamics, from
which we obtain the topology and functions, opposite to what is usually done in
the literature on Boolean Networks.

In fact, there exist many networks that display a given
trajectory. Even when the full state space structure is specified,
which defines the successor state of each of its $2^N$ possible
states, it is possible to construct a network that has this state
space structure. This can be done by constructing a fully connected graph
with $k=N$ and by assigning to each node the function that has the
required output for each of the $2^N$ input states. In the end, inputs
that never affect the output can be removed. If there are different
sets of inputs that can be simultaneously removed, different networks
are obtained.

When not the entire state space structure, but only one reliable trajectory is
specified, there exist consequently many networks with different topology and
functions which have this trajectory and may differ in the rest of their state
space.  We will restrict ourselves to \emph{minimal} networks, i.e., networks
with the smallest possible number of inputs for each node and the simplest
possible functions, which have the maximum possible number of identical entries
in the truth table.  This minimality condition is motivated by the putative cost
associated with more connections or more complicated functions, which would
decrease the fitness of an organism. This is in contrast to what was done
in~\cite{lau_function_2006}, where all possible networks were considered, which
is only feasible on very small systems.

Such minimal networks can be constructed by a straightforward
algorithm, because the inputs and the function required for each node
can be determined independently from those of all the other nodes.
The inputs for a given node must include all predecessor nodes, which
change their state 1 time step before the given node changes its
state.  Additional inputs are required if the given node assumes,
during the course of the trajectory, different binary states for the
same configuration of the predecessor nodes. The choice of these
``excess'' inputs is usually not unique and may include self
inputs. We perform this choice at random, but only from the
possibilities which minimize the number of inputs to each node. If not
all possible configurations of the states of the input nodes occur
during the course of the trajectory, the update function of the given
node is not unique. We first assign those truth table entries of the
update function that are specified by the trajectory. Then, we assign
to all remaining entries the same output value, and we choose the
majority of output values assigned so far. (If there is no majority,
we choose either value with probability 1/2.)

The algorithm used for choosing the minimal set of inputs proceeds as
follows: To each node, we first assign all predecessor nodes as
inputs. Then, if needed, we choose ``excess'' inputs. We first set the
number of excess inputs to $k'=1$, and we test in a random order the
${N \choose k'}$ possible node combinations until we find a node set
which, together with the predecessors, is a valid input set. If no
valid combination is found, we increase $k'$ by 1 and repeat the
search.  Once a valid combination is found, the corresponding truth
table is completed by applying the minimality condition to its
unspecified entries.  The run time of this algorithm increases as
$O(lN^{\max(\max(k'),1)})$, where $l$ is the average number of flips
per node, and $\max(k')$ is the maximum value of $k'$ for all
nodes. We have observed that the run times are feasible for networks
of size up to $N=400$ and $l=12$.~\footnote{We note that an 
  alternative procedure that starts with an input set that contains
  all nodes and then randomly removes inputs until a removal is no
  longer possible, is for larger $k$ much faster than the procedure used by
  us. However, it will in general not produce a minimal network. As an
  example, consider the case where only one minimal set of $k$ inputs
  is possible for a given node. In this case, if one of these inputs
  is removed early in the iteration, the input set obtained at the end
  will have a size larger than the minimal $k$.}

\begin{figure}[htb]
\includegraphics*[width=1.0\columnwidth]{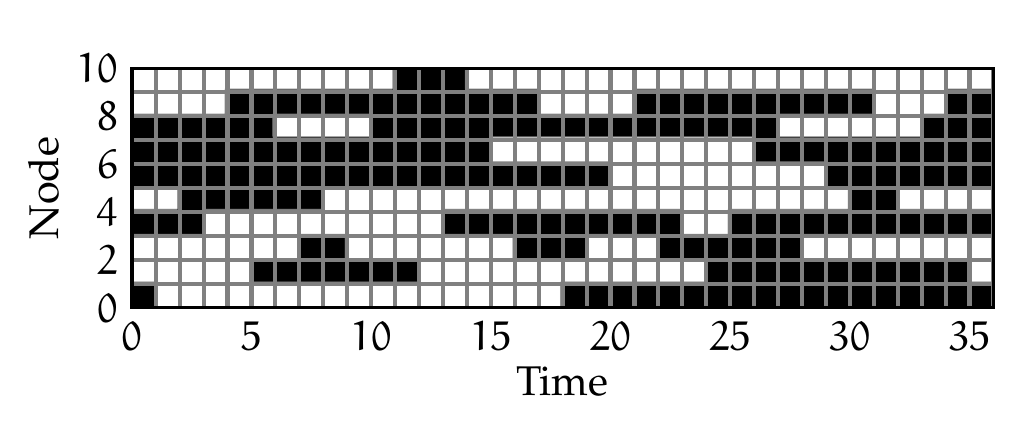}
\includegraphics*[width=0.45\columnwidth]{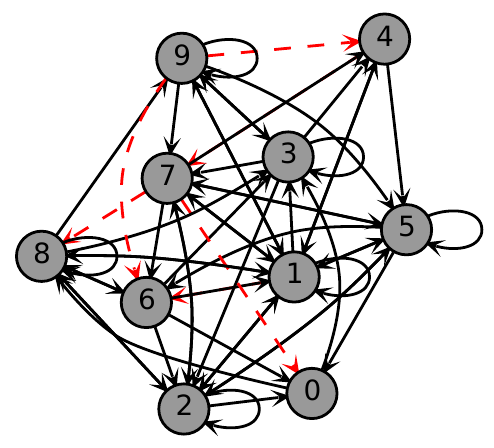} \hspace{0.5cm}
\includegraphics*[width=0.45\columnwidth]{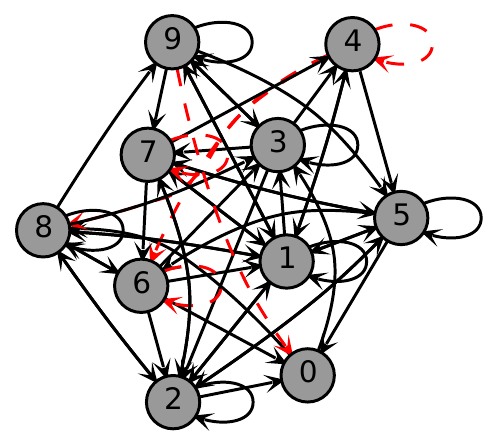}\\
\beginpgfgraphicnamed{fig-history-example}
\begin{tikzpicture}
\include{plots/functions1}
\end{tikzpicture}\hspace{0.5cm}
\begin{tikzpicture}
\include{plots/functions2}
\end{tikzpicture}
\endpgfgraphicnamed

\caption{(Color online) Example of a random reliable trajectory for $N=10$ and
  $l=4$, and two possible minimal networks. The edges with dashed (red) lines
  represent the inputs that are different between the two networks. Below each
  network are the outputs of the truth table of each node, ordered from top to
  bottom, and left to right, according to their input combination
  indices. Outputs marked in grey (cyan) correspond to input
  combinations present in the given trajectory.\label{fig:history_example}}
\end{figure}

An example for a reliable trajectory and two possible networks with
their functions, obtained with the above algorithm, is given in Figure
\ref{fig:history_example}.

We choose the reliable trajectory at random, without taking into consideration
possible particular features of biological networks, such as different temporal
activation patterns of the different nodes, which reflect the function that the
network must fulfill. Instead, we will consider a null model, where the values
of the nodes change randomly. The only restriction which is imposed is that the
trajectory is reliable. The only two parameters of this trajectory are the
number of nodes $N$, and the average number of flips per node $l$. We generate a
random ensemble of reliable trajectories in the following way: First, we
determine how often each node shall be flipped. To this purpose, for each node
$i$ a random number $\lambda_i$ is chosen from a Poisson distribution with mean
$(l-2)/2$, implying that node $i$ shall be flipped $l_i\equiv 2\lambda_i+2$
times. The average number of flips of each node is thus identical to $l$, and
each node is flipped at least twice. The length of the trajectory is then
$L=2N+2\sum_i \lambda_i=\sum_i l_i$. Then, we arrange these flips in a randomly
chosen order. If the resulting trajectory contains the same network state twice,
it is discarded, and a new sequence of flips is chosen.

\subsection{Topological characteristics}

We first present results for the topological characteristics of the obtained
networks. We evaluate the degree distribution and the local correlations. The
degree distribution is of course strongly dependent on $l$.  Local correlations
can arise when two nodes that are influenced by the same nodes are more likely
to influence each other.

Unless stated otherwise, we averaged the results from several independent
realizations of the minimal trajectories and minimal networks, for different $N$
and $l$. The number of realizations for small $N$, up to $20$, were at least
$2000$. For intermediary values of $N$, up to $100$, it varied from 50 to 300,
depending on $l$. For the larger networks, $N>100$, it ranged from 200 to 6
networks for $l<12$, and one realization for $N=400$ and $l=12$.

\subsubsection{Degree distribution}

\begin{figure}[htb]
\includegraphics*[width=1.0\columnwidth]{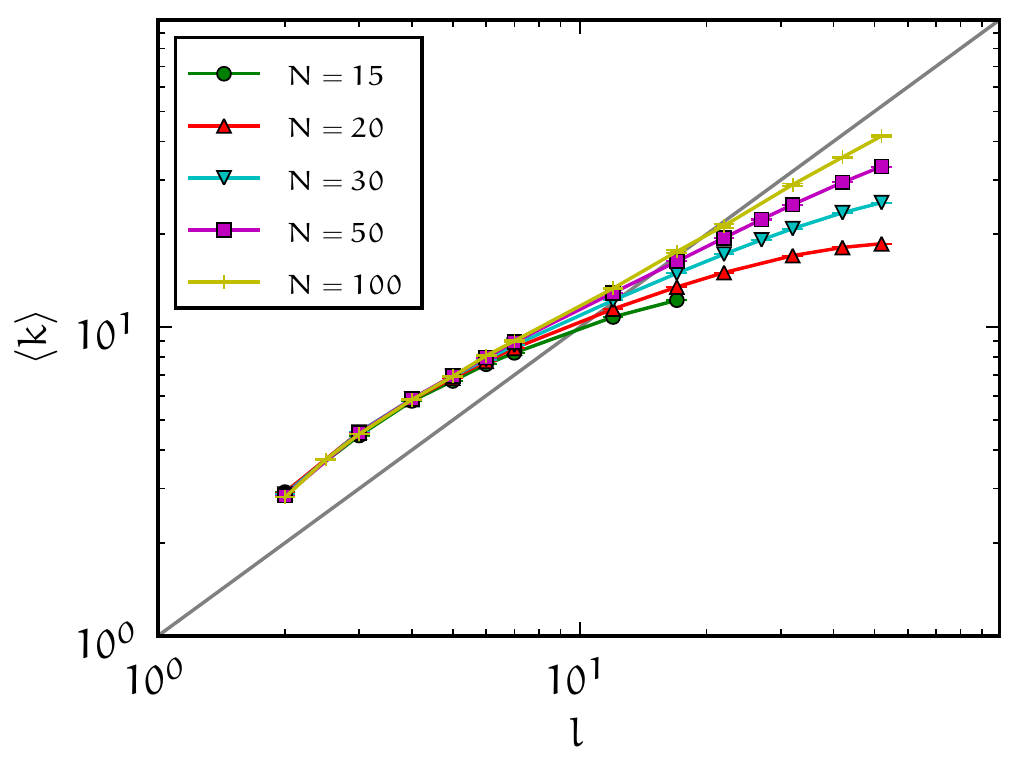}
\caption{(Color online) Average degree $\avg{k}$ as function of $l$ for
  networks of different size $N$. The straight line is the function
  $\avg{k}=l$.\label{fig:avgk-vs-l}}
\end{figure}

The number of inputs of a node is at least as large as the number of its
predecessors. Whenever the state of the node cannot be written as a function of
the predecessors alone, ``excess'' inputs must be chosen, as already mentioned
before. The number of different predecessors $n_p$ per node approaches, for
large $N$, on average $l$, since it becomes unlikely for large $N$ that the same
node is chosen twice as predecessor.  The typical truth table size grows
therefore with $l$ as $ 2^{l}$. Since the number of different predecessor states
grows only quadratically as $n_pl\sim l^2$, one can expect the number of
``excess'' inputs to be small, and number of inputs per node should be
\begin{equation}
\avg{k} \simeq l,
\end{equation}
for sufficiently large $N$. This is confirmed by our numerical
investigations, as is shown in Fig.~\ref{fig:avgk-vs-l}.

\FloatBarrier

\begin{figure}[htb]
\includegraphics*[width=1.0\columnwidth]{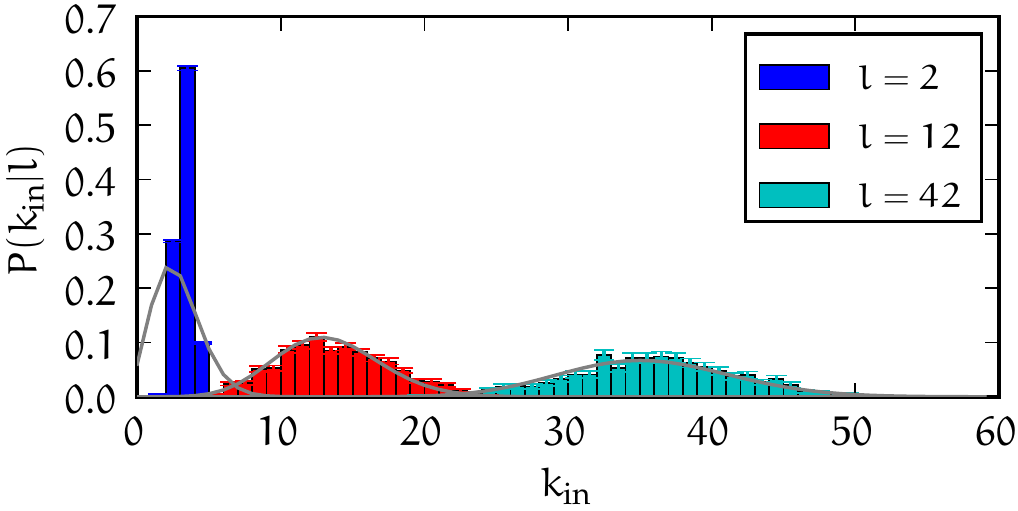}
\includegraphics*[width=1.0\columnwidth]{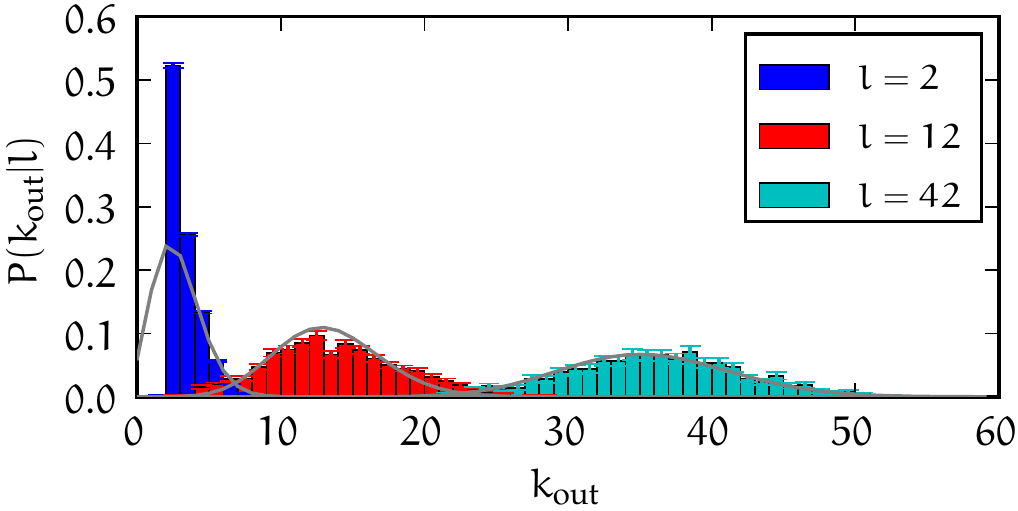}
\caption{(Color online) In-degree and out-degree distributions of minimal
  networks for different values of $l$, for $N=100$. The solid lines correspond
  to Poisson distributions with the same average. \label{fig:deg_dist}}
\end{figure}

The degree distribution mirrors the distribution of the number of
predecessors. Since all nodes flip on average the same number of times, the
distribution is expected to follow a Poisson distribution for large enough
$l$. This is indeed the case, as Fig.~\ref{fig:deg_dist} shows. For small $l$
however, the distributions are more narrow, because we imposed the condition
that each node flips at least twice, leaving little freedom for additional
predecessors when $l$ is close to 2.

\subsubsection{Local correlations}

\begin{figure}[htb]
\includegraphics*[width=0.45\columnwidth]{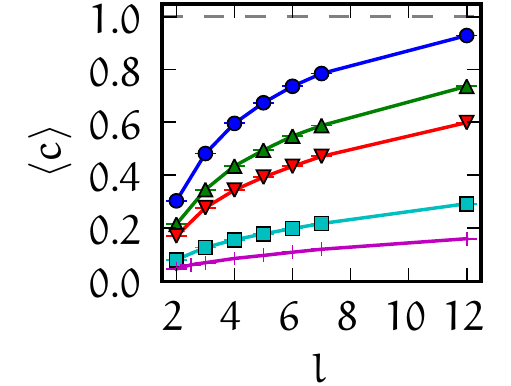}
\includegraphics*[width=0.45\columnwidth]{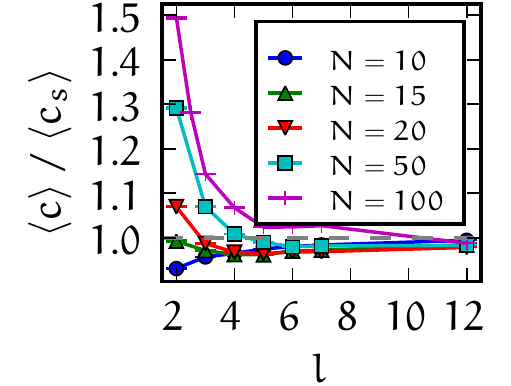}
\includegraphics*[width=0.45\columnwidth]{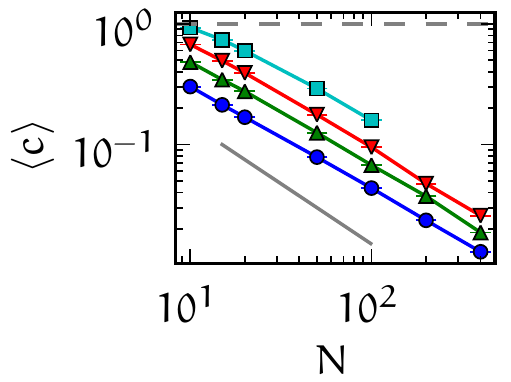}
\includegraphics*[width=0.45\columnwidth]{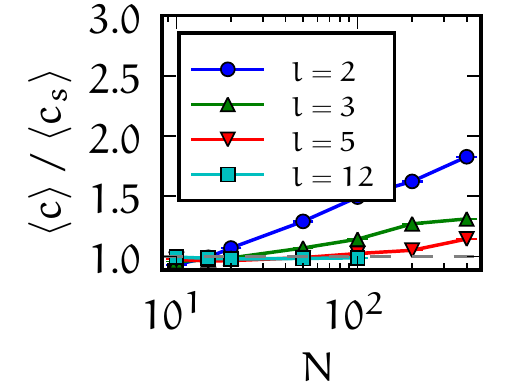}
\caption{(Color online) Clustering ratio $\avg{c}/\avg{c_s}$ as a function of
  the average number of flips per node $l$, for different network sizes $N$. The
  gray straight line corresponds to a decay of the type
  $1/N$. \label{fig:clust}}
\end{figure}

We obtained information about the local topology of the minimal
networks by evaluating the probability that the neighbours of a given
node are connected to each other. This probability is the so-called
\emph{clustering coefficient} $\avg{c}$ \cite{newman_structure_2003}.
Random uncorrelated networks show absence of clustering only in the
limit $N\to\infty$. Thus, for finite $N$, it is necessary to compare
the obtained value with a random network of equal size and with equal
degree distribution. In order to do this, we calculated the clustering
coefficient $\avg{c_s}$ on \emph{shuffled} networks, where the links
were rewired randomly, preserving the in- and out-degree of each node.
We then calculated the ratio $\avg{c}/\avg{c_s}$, for networks of
different size and average flip number $l$. If the ratio approaches
$1$, the network does not exhibit any special clustering.  The results
for several values of $N$ and $l$ are shown in Fig.~\ref{fig:clust}.

The most evident feature of Fig.~\ref{fig:clust} is that clustering is stronger
for smaller $l$, i.e., for sparse networks. For larger $l$ (and hence larger
$\avg{k}$), the average distance between nodes decreases, and the shuffled and
original networks have a similar degree of clustering. This difference between
networks with smaller and larger average degree becomes more pronounced when the
size of the networks $N$ is increased. From the data in Fig.~\ref{fig:clust}, it
appears that that the ratio $\avg{c}/\avg{c_s}$ increases slowly with $N$. We
will argue in the following that this ratio will reach a finite asymptotic value
in the limit $N\to \infty$.

\begin{figure}[bht]
\includegraphics*[width=1.0\columnwidth]{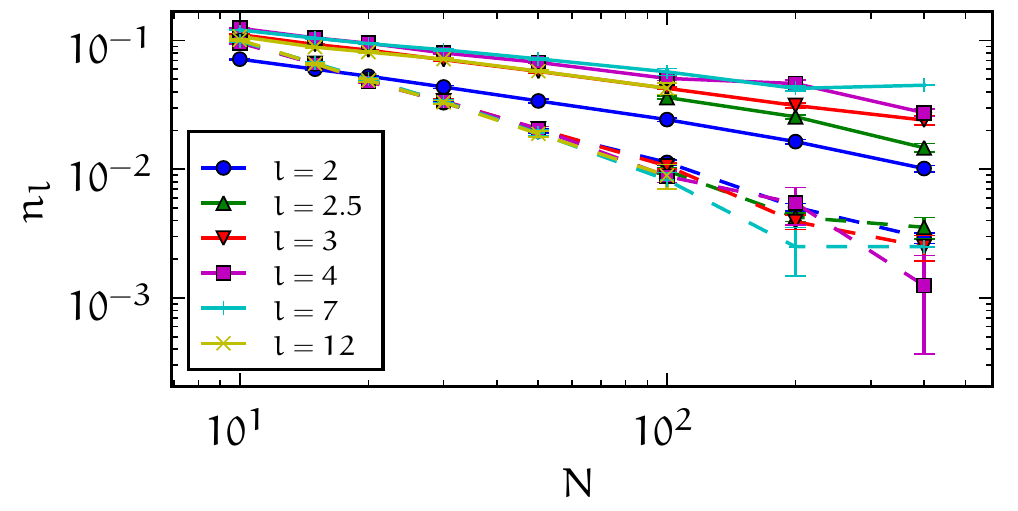}
\caption{(Color online) Fraction $n_l$ of nodes with self-input, as a function
  of $N$, for different values of $l$.  The dashed curves are obtained for
  shuffled networks, with the same degree sequence. The inset shows the ratio
  $n_l/n^s_l$, where $n^s_l$ is the self-input ratio for the shuffled
  networks. \label{fig:loop-vs-n}}
\end{figure}

The finding that the clustering coefficients are larger than for random networks
can be explained by considering the above-mentioned excess inputs that are
required when the function assigned to a node cannot be based on its predecessor
nodes alone. Let us consider two consecutive flips of a node $j$ on a given
trajectory. These flips are preceded by flips of the predecessor nodes, which we
call $v$ and $w$. The average time between the two considered flips of node $j$
is $\sim L/l = N$, implying that there is a considerable probability that node
$v$ flips again before the second flip of node $j$, giving the sequence
  \begin{equation*}
    vj \cdots v \cdots wj \, .
  \end{equation*}
The update function assigned to node $j$ needs an excess input if neither node
$w$ nor any other predecessor of node $j$ (which can exist only for $l>2$)
flips between the first flip of $j$ and the second flip of $v$.  The simplest
choice of this excess input is node $j$ itself. Indeed, self-inputs occur more
often than in the shuffled networks, as is shown in
Fig.~\ref{fig:loop-vs-n}. Since the number of different possible excess inputs
is proportional to $N$, we expect that the fraction $n_l$ of nodes with
self-inputs decreases as $n_l\sim 1/N$ for large $N$, but remains larger than
that of shuffled networks by a constant multiplicative factor.

The excess input cannot be a self-input if node $w$ flips also in the same
interval, giving the sequence
 \begin{equation*}
    vj \cdots v \cdots w\cdots wj \, .
  \end{equation*}
In this case, an excess input $u$ must be chosen among those nodes that flip
between the two consecutive flips of node $w$, if none of the other predecessors
of $j$ flips in this interval, giving the sequence
 \begin{equation*}
    vj \cdots v \cdots w\cdots u\cdots wj \, .
  \end{equation*}
Now, the average distance between the flips of node $w$ and node $u$ is smaller
than that between two randomly chosen nodes, since $w$ is required to flip in
the indicated interval. Therefore, the probability that $w$ is an input to $u$
or vice versa is larger than random, and it scales as $1/N$ in the limit $N \to
\infty$. Since $w$ and $u$ are inputs to $j$, it follows that the clustering
coefficient is larger than the random value $\avg{c_s}$.

From this consideration, it follows that the ratio $\avg{c}/\avg{c_s}$
approaches a constant value in the limit $N\to \infty$. Furthermore, it follows
that this ratio is larger for smaller $l$, since it is less likely that there
exist additional inputs to $j$ that flip in the required interval and make
excess inputs unnecessary. The slight increase seen in
Fig.~\ref{fig:clust} can probably be attributed to a finite-size effect.

\begin{figure}[hbt]
  \includegraphics*[width=1\columnwidth]{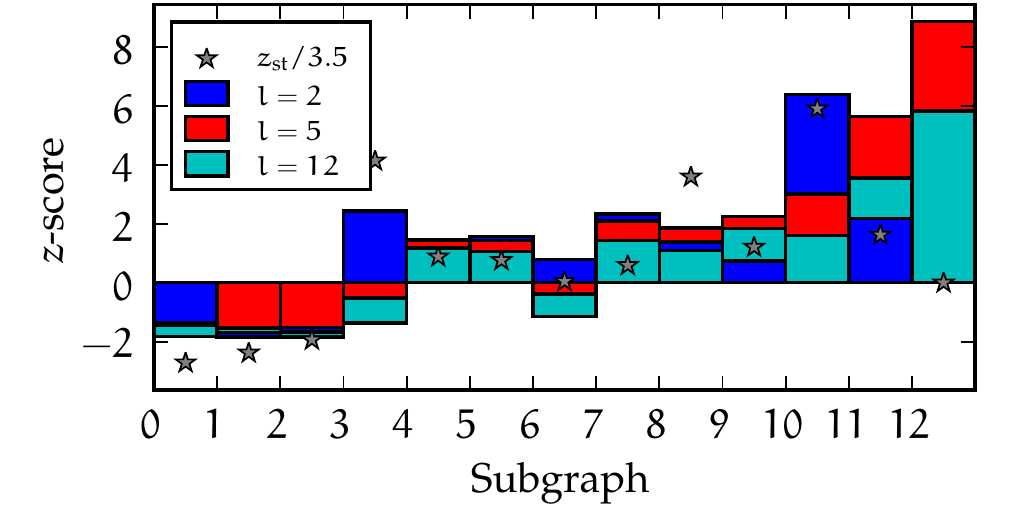}
  \beginpgfgraphicnamed{fig-motifs}
  \begin{minipage}{0.9\columnwidth}
    \include{plots/motifs-s3-n100-no-loops-foo}
  \end{minipage}
  \endpgfgraphicnamed
\caption{(Color online) The $z$-score of the different three-node subgraphs of
  minimal reliable networks, for different values of $l$ and $N=100$. The
  profile $z_\text{st}$ corresponds to the the signal-transduction interaction
  network in mammalian cells~\cite{milo_superfamilies_2004}. \label{fig:motifs}}
\end{figure}
\FloatBarrier

In order to determine which three-node subgraphs contribute to the
increased clustering, we evaluated their $z$-score, which indicates to
what extent the frequency of each subgraph is different compared to
the random case. The $z$-score is defined as
\begin{equation}
  z_i = \frac{\avg{N_i} - \avg{N^s_i}}{\sqrt{\avg{(N^s_i)^2} - \avg{N^s_i}^2}},
\end{equation}
where $N_i$ is the number of occurrences of subgraph $i$, and $N_i^s$ is the
number of occurrences of the same subgraph on a shuffled network with the same
degree sequence. Fig.~\ref{fig:motifs} shows the different possible subgraphs
and their $z$-score. Subgraphs with more links have a higher $z$-score and are
therefore \emph{network motifs}. Sparser subgraphs, where there is no link
between two of the nodes, are rarer than at random, as predicted by the
clustering coefficient.  The abundance of denser motifs increases with $l$, as
the network itself becomes more dense, but the overall trend of the $z$-score is
the same. One peculiar feature is the absence of simple loops (subgraph 6), also
know as feedback loops~\cite{alon_introduction_2007}. As was described above,
the clustering is mostly due to the correlations between the inputs of a given
node. A simple loop does not have this type of correlation. Furthermore, it was
shown by Klemm et al~\cite{klemm_topology_2005} in a study of the reliability of
small Boolean networks, that feedback loops are harmful to reliable
dynamics. These authors obtained a $z$-score profile very similar to
Fig.~\ref{fig:motifs} (see Fig.~4 of~\cite{klemm_topology_2005}). They also
showed that this profile is qualitatively similar to real
biological networks studied in~\cite{milo_superfamilies_2004}. A direct
comparison is shown in Fig.~\ref{fig:motifs}, with the motif profile of the
signal-transduction interaction network in mammalian
cells~\cite{milo_superfamilies_2004}.

\begin{figure}[htb]
  \subfloat[$k=2$]{\includegraphics*[width=1.0\columnwidth]{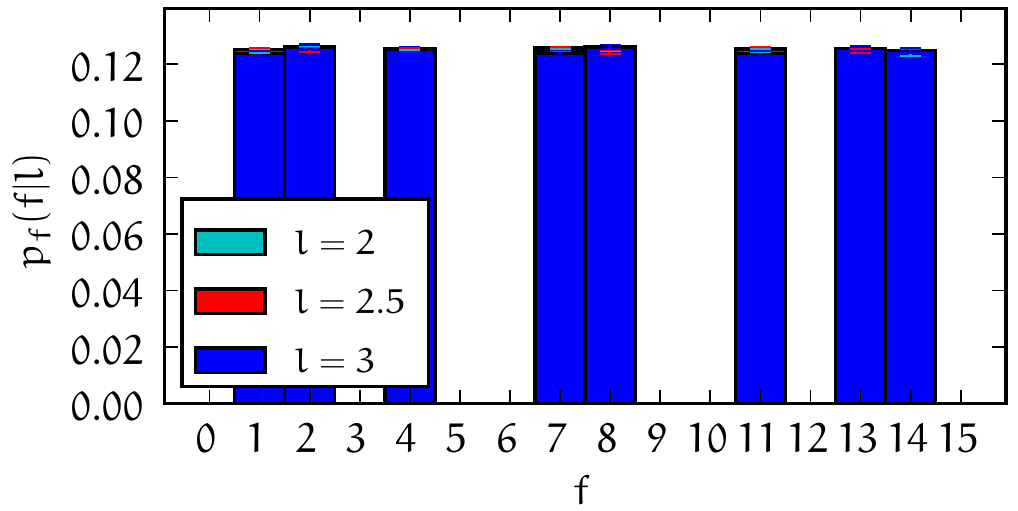}}\\
  \subfloat[$k=3$]{\includegraphics*[width=1.0\columnwidth]{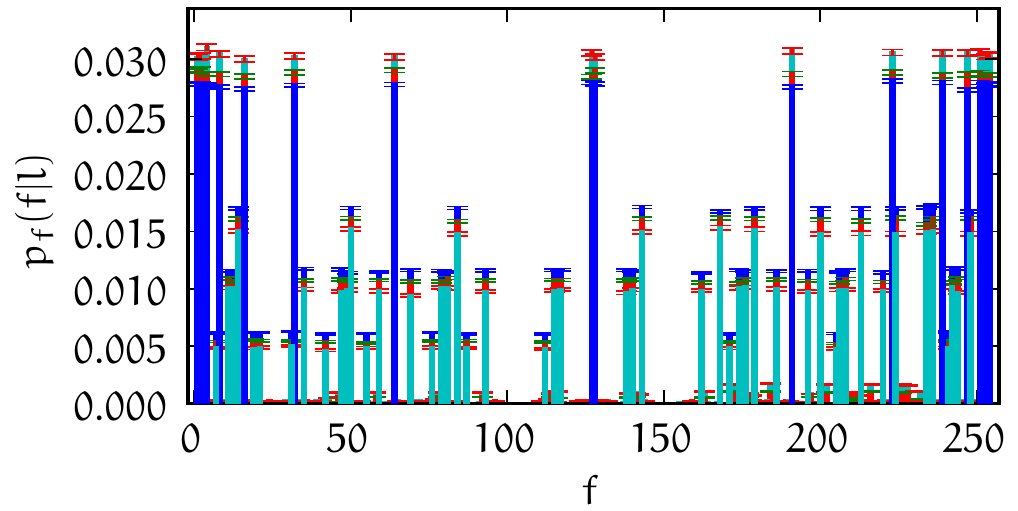}}\\
  \subfloat[$k=3$, without self-loops.]
           {\label{fig:fhist-k3-nl}
             \includegraphics*[width=1.0\columnwidth]{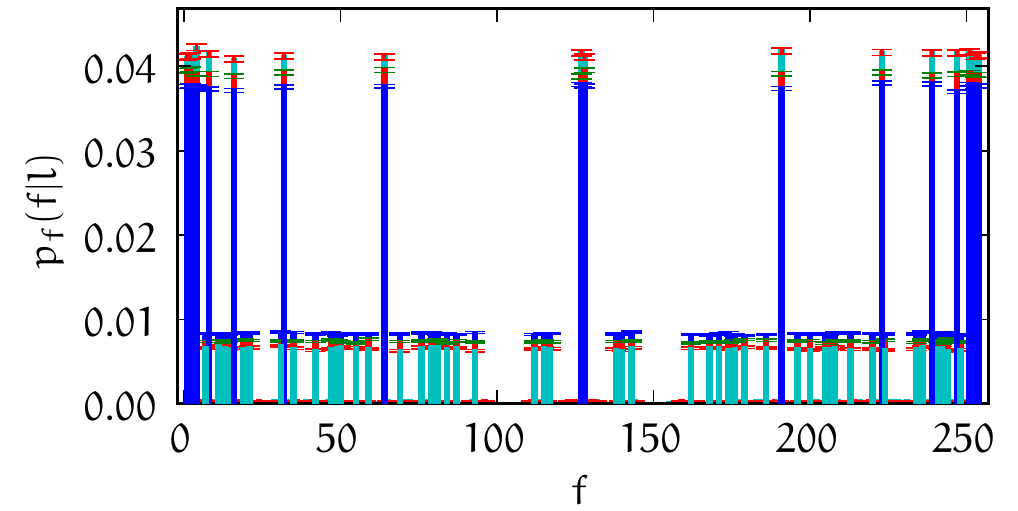}}\\
  \subfloat[$k=4$, without self-loops.]
           {\includegraphics*[width=1.0\columnwidth]{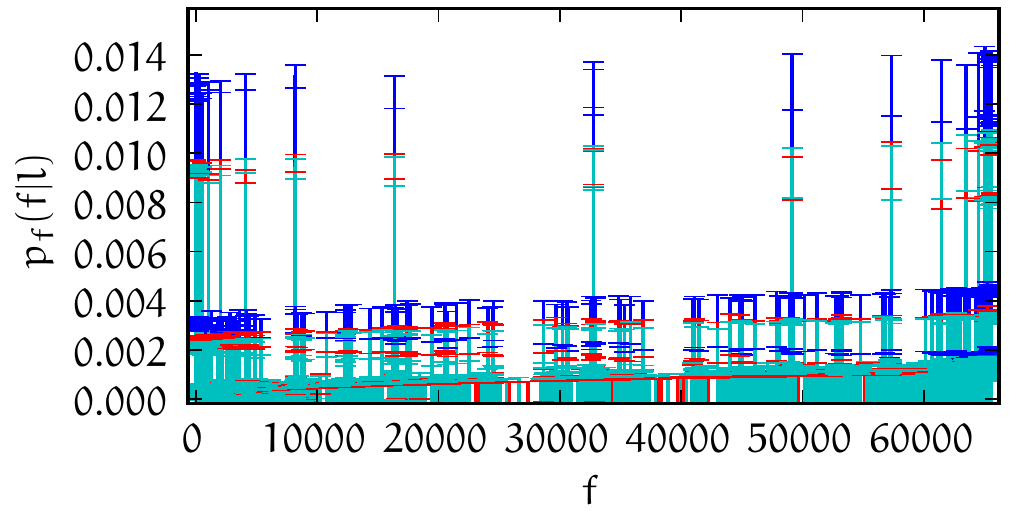}}\\
  \caption{(Color online) Distribution of the different update
    functions, for different numbers of inputs, $k$, and for different
    flip numbers, $l$, for networks of size $N=20$. \label{fig:func-dist}}
\end{figure}
\FloatBarrier

In Fig~\ref{fig:motifs}, we did not keep track of the self-inputs, for
simplicity. When self-loops are included in the subgraphs, their
number increases from 13 to 86, which makes the analysis and
presentation more elaborate. We performed this analysis and found that
a subgraph with a specific number of self-loops has a larger $z$-score
than its counterpart with less or no self-loops. The $z$-score pattern
of Fig.~\ref{fig:motifs}, on the other hand, is repeated for
subgraphs which share the same number of self-loops, which shows that
motif occurrence and self-regulation are largely independent.

\subsection{Properties of update functions}\label{sec:functions}

We evaluated the frequency of the different types of update functions
in minimal networks, for different values of $l$, see
Fig.~\ref{fig:func-dist}. Unless otherwise stated, the results were
obtained from $10^4$ independent realizations of networks with
$N=20$. We compared the results with those obtained for larger values
of $N$, with no discernible difference other than the reduced
statistical quality. Functions with different numbers of inputs were
evaluated separately.

The functions seem to be distributed according to different classes, where
functions of the same type occur with the same probability, while some do not
occur at all. In order to understand this distribution, it is necessary to
describe in detail what conditions need to be met by the functions, according to
the imposed dynamics and construction rules.

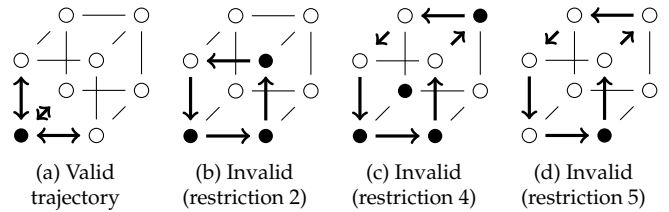
\begin{figure}[htb]
  \newcommand{\bit}[1]{
    \ifnum #1<1
    \tikz \fill[black] (0,0) circle (0.3em);
    \else
    \tikz \draw (0,0) circle (0.3em);
  \fi
  }
  \subfloat[Valid \mbox{trajectory}]{\label{fig:traj_valid}
    \beginpgfgraphicnamed{fig-traj-valid}
    \begin{tikzpicture}
      \node (000) at (0,0) [] {\bit{0}};
      \node (001) at (1.0,0) [] {\bit{1}} edge [<->,very thick] (000);
      \node (010) at (0,1.0) [] {\bit{1}} edge [<->,very thick] (000);
      \node (011) at (1.0,1.0) [] {\bit{1}} edge [-] (010) edge [-] (001);

      \node (100) at (0.6,0.6) [] {\bit{1}} edge [<->,very thick] (000);
      \node (101) at (1.6,0.6) [] {\bit{1}} edge [-] (100) edge [-] (001);
      \node (110) at (0.6,1.6) [] {\bit{1}} edge [-] (010) edge [-] (100);
      \node (111) at (1.6,1.6) [] {\bit{1}} edge [-] (110) edge [-] (011) edge [-] (101);
    \end{tikzpicture}
    \endpgfgraphicnamed
  }
  \subfloat[Invalid \mbox{(restriction~2)}]{\label{fig:traj_r2}
    \beginpgfgraphicnamed{fig-traj-r2}
    \begin{tikzpicture}
      \node (000) at (0,0) [] {\bit{0}};
      \node (001) at (1.0,0) [] {\bit{0}} edge [<-,very thick] (000);
      \node (010) at (0,1.0) [] {\bit{1}} edge [->,very thick] (000);
      \node (011) at (1.0,1.0) [] {\bit{0}} edge [->, very thick] (010) edge [<-,very thick] (001);

      \node (100) at (0.6,0.6) [] {\bit{1}} edge [-] (000);
      \node (101) at (1.6,0.6) [] {\bit{1}} edge [-] (100) edge [-] (001);
      \node (110) at (0.6,1.6) [] {\bit{1}} edge [-] (010) edge [-] (100);
      \node (111) at (1.6,1.6) [] {\bit{1}} edge [-] (110) edge [-] (011) edge [-] (101);
    \end{tikzpicture}
    \endpgfgraphicnamed
  }
  \subfloat[Invalid \mbox{(restriction~4)}]{\label{fig:traj_r3}
    \beginpgfgraphicnamed{fig-traj-r3}
    \begin{tikzpicture}
      \node (000) at (0,0) [] {\bit{0}};
      \node (001) at (1.0,0) [] {\bit{0}} edge [<-,very thick] (000);
      \node (010) at (0,1.0) [] {\bit{1}} edge [->,very thick] (000);
      \node (011) at (1.0,1.0) [] {\bit{1}} edge [-] (010) edge [<-,very thick] (001);

      \node (100) at (0.6,0.6) [] {\bit{0}} edge [-] (000);
      \node (101) at (1.6,0.6) [] {\bit{1}} edge [-] (100) edge [-] (001);
      \node (110) at (0.6,1.6) [] {\bit{1}} edge [->, very thick] (010) edge [-] (100);
      \node (111) at (1.6,1.6) [] {\bit{0}} edge [->, very thick] (110) edge [<-,very thick] (011) edge [-] (101);
    \end{tikzpicture}
    \endpgfgraphicnamed
  }
  \subfloat[Invalid \mbox{(restriction~5)}]{\label{fig:traj_r4}
    \beginpgfgraphicnamed{fig-traj-r4}
    \begin{tikzpicture}
      \node (000) at (0,0) [] {\bit{1}};
      \node (001) at (1.0,0) [] {\bit{0}} edge [<-,very thick] (000);
      \node (010) at (0,1.0) [] {\bit{1}} edge [->,very thick] (000);
      \node (011) at (1.0,1.0) [] {\bit{1}} edge [-] (010) edge [<-,very thick] (001);

      \node (100) at (0.6,0.6) [] {\bit{1}} edge [-] (000);
      \node (101) at (1.6,0.6) [] {\bit{1}} edge [-] (100) edge [-] (001);
      \node (110) at (0.6,1.6) [] {\bit{1}} edge [->, very thick] (010) edge [-] (100);
      \node (111) at (1.6,1.6) [] {\bit{1}} edge [->, very thick] (110) edge [<-,very thick] (011) edge [-] (101);
    \end{tikzpicture}
    \endpgfgraphicnamed
  }
  \hfill {}
  \caption{Example of input and output trajectories on the $k$-hypercube
    representing the states of the inputs, for functions with $k=3$. Allowed
    transitions are represented by arrows. The color on each vertex represent
    the output value. Fig. (a) represents one type of valid
    trajectory. Figs.~\subref{fig:traj_r2} to \subref{fig:traj_r4} represent
    invalid trajectories according to the indicated restriction:
    \subref{fig:traj_r2} not all sides of the cube are visited;
    \subref{fig:traj_r3} the function is not minimal; \subref{fig:traj_r4} the
    function can be reduced to $k=2$.\label{fig:traj-types}}
\end{figure}

The subsystem composed only of the inputs of a given node follows a
certain ``local trajectory'' (i.e., sequence of states), which
determines, together the minimality condition described in
Sec.~\ref{sec:minimal}, the update function of the considered
node. The probabilities of the different possible trajectories depend
on the way the global trajectory is specified, and on the rules for
choosing excess inputs. The restrictions imposed on the local
trajectories of the inputs are as follows:
\begin{enumerate}
  \item The local trajectory of the inputs must correspond to a
    periodic walk on the $k$-dimensional hypercube representing their
    states, since the Hamming distance at each step must be $1$. We
    note that in this subsystem, the same input state is allowed to
    repeat within a period (only the global state cannot). The
    vertices of the hypercube can be annotated with the output value
    of the function at the corresponding input state (see
    Fig.~\ref{fig:traj-types} for examples).
  \item For large $N$, the trajectories of any two different nodes
    will be approximately random and uncorrelated. The only
    restriction is that every face of the hypercube will be visited
    exactly $l_v$ times, where $v$ is the index of the input node that
    has a fixed state on this face. On
    average we have $\avg{l_v}=l$.
  \item The output values of the function can be distributed on the
    vertices of the hypercube that are visited during the walk in any
    possible way, with the restriction that the output value must
    change $l_j$ times along the walk, where $j$ is the index of the
    considered node.  An exception are functions with
    self-inputs: the vertices on the hypercube face corresponding to
    the self-input must all have the same output value.
  \item The output values at the vertices of the hypercube which
    are not visited by the walk must be equal to the majority of the output
    values on the walk (this is the minimality condition defined in
    section~\ref{sec:minimal}).
  \item Functions that can be reduced to a function with smaller $k$
    cannot occur due to the minimality condition, and the
    corresponding trajectory can be confined to a
    hypercube of smaller dimension.
\end{enumerate}
Fig.~\ref{fig:traj-types} shows examples of trajectories that are allowed or not
allowed for the case $k=3$.

The listed restrictions result in the observed distribution of update functions.
We will describe in detail all the possibilities for $k=2$, and discuss in a
more general and approximate manner the functions with $k>2$.

\subsubsection{Functions with $k=2$}

\begin{figure}[htb]
  \beginpgfgraphicnamed{fig-traj-k2}
  \newcommand{\bit}[1]{
    \ifnum #1<1
    \tikz \fill[black] (0,0) circle (0.25em);
    \else
    \tikz \draw (0,0) circle (0.25em);
  \fi
  }
  \newcommand{\func}[4]{
    \begin{tikzpicture}
        \node (00) at (0,0) [] {\bit{#1}};
        \node (01) at (0,0.5) [] {\bit{#2}} edge [-] (00);
        \node (10) at (0.5,0) [] {\bit{#3}} edge [-] (00);
        \node (11) at (0.5,0.5) [] {\bit{#4}} edge [-] (10) edge [-] (01);
    \end{tikzpicture}
  }

  \begin{tikzpicture}
    \node at (0,0) {\func{0}{0}{0}{0}}; \node at (0,1) {\func{1}{1}{1}{1}};
    \node at (1,0) {\func{1}{1}{0}{0}}; \node at (1,1) {\func{0}{0}{1}{1}};
    \node at (2,0) {\func{0}{1}{0}{1}}; \node at (2,1) {\func{1}{0}{1}{0}};
    \node at (3,0) {\func{1}{0}{0}{1}}; \node at (3,1) {\func{0}{1}{1}{0}};

    \draw (3.75, -0.25) -- (3.75, 1.25);

    \node at (4.5,0) {\func{1}{0}{0}{0}}; \node at (4.5,1) {\func{0}{1}{1}{1}};
    \node at (5.5,0) {\func{0}{1}{0}{0}}; \node at (5.5,1) {\func{1}{0}{1}{1}};
    \node at (6.5,0) {\func{0}{0}{0}{1}}; \node at (6.5,1) {\func{1}{1}{1}{0}};
    \node at (7.5,0) {\func{0}{0}{1}{0}}; \node at (7.5,1) {\func{1}{1}{0}{1}};
  \end{tikzpicture}
  \endpgfgraphicnamed
  \caption{Representation of all 16 functions with $k=2$ on the
    2-hypercube. On the left are the functions which do not (or
    rarely) occur in the minimal networks, and on the right are the
    canalysing functions which occur with equal
    probability.\label{fig:traj-k2}}
\end{figure}
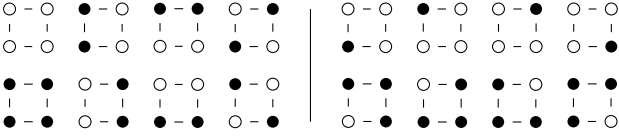

Fig.~\ref{fig:func-dist} shows that only 8 of the 16 possible
functions occur, and all of them with equal probability.  They are all
\emph{canalyzing functions}, with three entries 1 (or 0) in the truth
table, and one entry 0 (or 1). The hypercube representation of all
functions is shown in Fig.~\ref{fig:traj-k2}. The functions that are
not possible are obviously the constant functions (first row of
Fig.~\ref{fig:traj-k2}, from left to right), and the functions which
are insensitive to one of their inputs, due to restriction 4 (second and
third row). The other functions which do not occur are the reversible
functions, which change the output at every change of an input (fourth
row). Those functions, however, are not entirely impossible: It is
possible to construct a trajectory that meets all the listed
requirements, with the specification that the output flips as often
as all inputs together (restrictions 2 and 3). Such
trajectories follow the pattern
\begin{equation*}
  vj\cdots wj \cdots vj\cdots wj,
\end{equation*}
where $v$ and $w$ are the inputs of $j$.  This pattern is impossible
for $l=2$, but can occur for larger $l$, albeit with a small
probability, since $k=2$ functions are less likely for larger $l$;
furthermore, the probability that a node has two predecessors which
occur twice decreases with $N$ as $\sim1/N^2$.

\subsubsection{Functions with $k>2$}

\begin{figure}[htb]
  \subfloat[All functions]
           {\label{fig:all-classes}\includegraphics*[width=0.49\columnwidth]{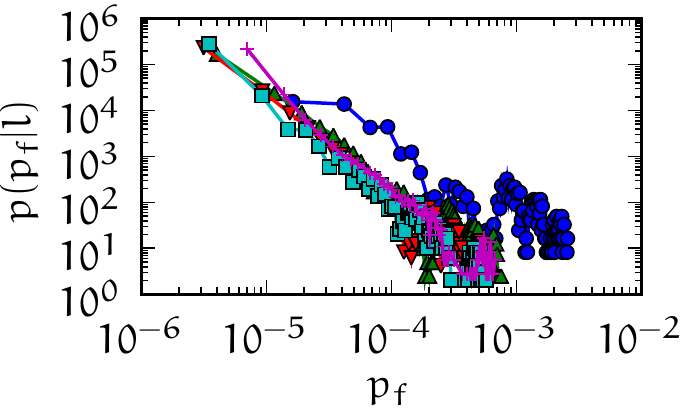}}
  \subfloat[No self-loops]
           {\label{fig:all-classes-nl}\includegraphics*[width=0.49\columnwidth]{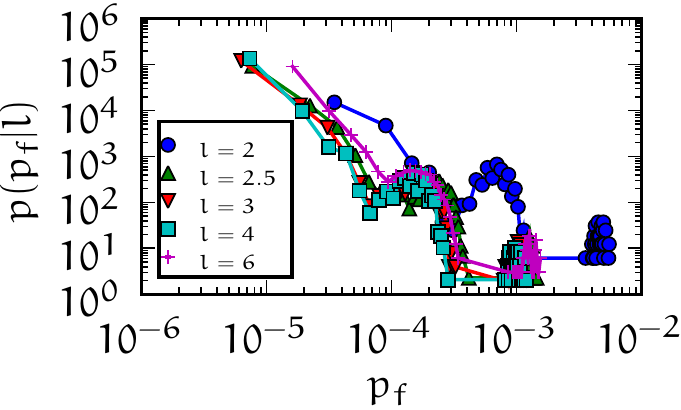}}\\
  \subfloat[$d=1$, no self-loops]
           {\label{fig:class-d1}\includegraphics*[width=0.49\columnwidth]{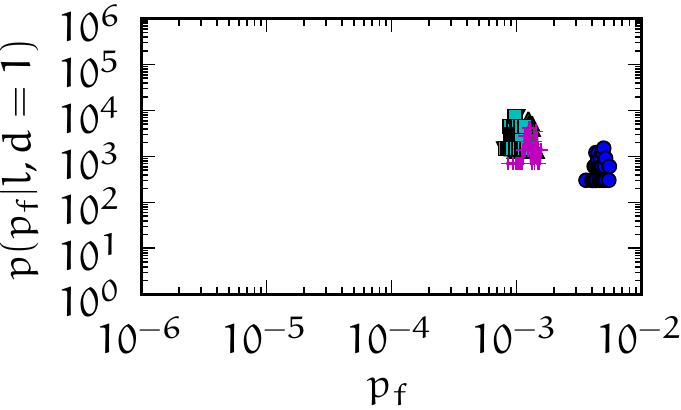}}
  \subfloat[$d=2$, no self-loops]
           {\label{fig:class-d2}\includegraphics*[width=0.49\columnwidth]{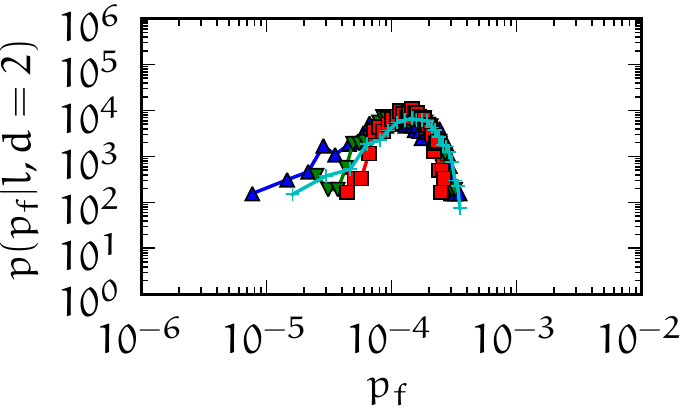}}\\
  \subfloat[$d=3$, no self-loops]
           {\label{fig:class-d3}\includegraphics*[width=0.49\columnwidth]{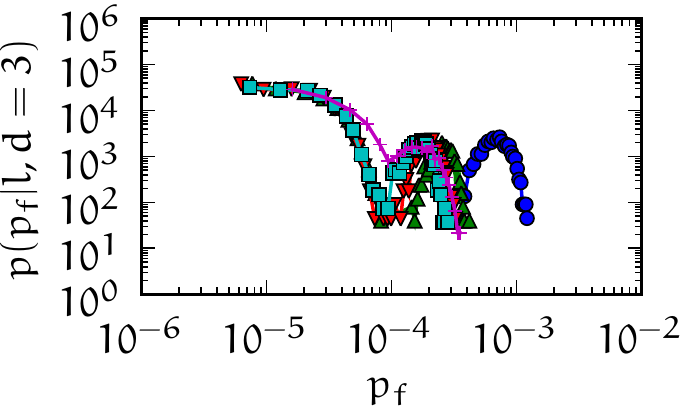}}
  \subfloat[$d=4$, no self-loops]
           {\label{fig:class-d4}\includegraphics*[width=0.49\columnwidth]{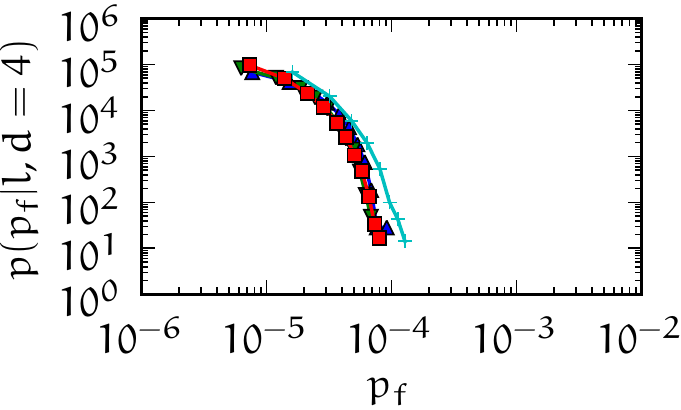}}
  \caption{(Color online) Distribution of function weights $p_f$, subdivided
    according to the value of the truth table homogeneity $d$, for different
    values of the average flip number $l$, for fixed values $k=5$ and
    $N=20$. \label{fig:d-classes}}
\end{figure}

Functions with $k>2$ seem to fall into different classes, which occur
with different probabilities. This can be seen by plotting the
distribution of the probabilities $p_f$ of the different functions, as
shown in Fig.~\ref{fig:d-classes}\subref{fig:all-classes} for
$k=5$. The different classes seem to correspond to different function
homogeneity values, defined as the number of minority output values
in the truth table, $d$. This can be verified by selecting only those
functions with a given value of $d$, and plotting their distribution
of probabilities, as shown in
Figs~\ref{fig:d-classes}\subref{fig:class-d1}
to~\subref{fig:class-d4}.  The most frequent class comprises the
functions with only one entry in the truth table deviating from the
others ($f=2^i$ and $f=2^k - 2^i$), with $d=1$ (see
Fig~\ref{fig:d-classes}\subref{fig:class-d1}). Those are
\emph{canalysing} functions, where all inputs are canalysing
inputs. Functions with the same homogeneity fall into subclasses which
have different probabilities. Those functions are often negated
functions ($f' = 2^k-f$) of one another, and this is due to the
existence of self-inputs: Self-regulated functions are not equivalent
functionally when they are negated (the input corresponding its own
output must be negated as well), despite sharing the same
homogeneity. The $0 \leftrightarrow 1$ symmetry, however, is always
preserved.  When self-loops are ignored, the distribution becomes
symmetric with respect to negation of the output (see
Fig~\ref{fig:func-dist}\subref{fig:fhist-k3-nl}), and the homogeneity
classification becomes the predominant criterion to distinguish between
the classes (compare Figs~\ref{fig:d-classes}\subref{fig:all-classes}
and~\subref{fig:all-classes-nl}). But even in the absence of
self-loops, the probability classes are not uniquely defined by the
homogeneity, and there are overlaps between the different classes, as
Figs~\ref{fig:d-classes}\subref{fig:class-d2} to~\subref{fig:class-d4}
show. Nevertheless, there is a general tendency that functions with
larger $d$ are less likely.

\begin{figure}[htb]
  \subfloat[$k=3$]{\includegraphics*[width=0.49\columnwidth]{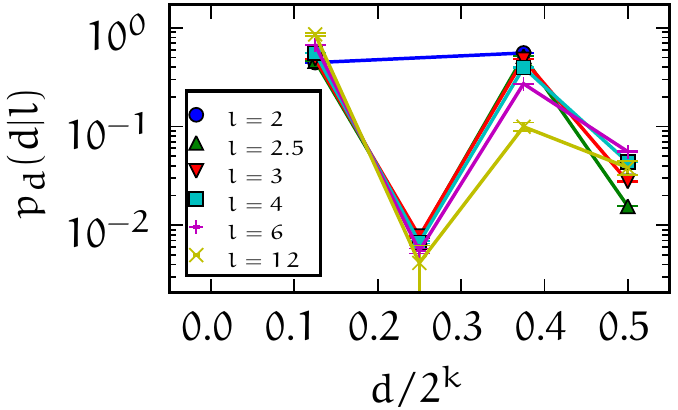}\hfill
                   \includegraphics*[width=0.49\columnwidth]{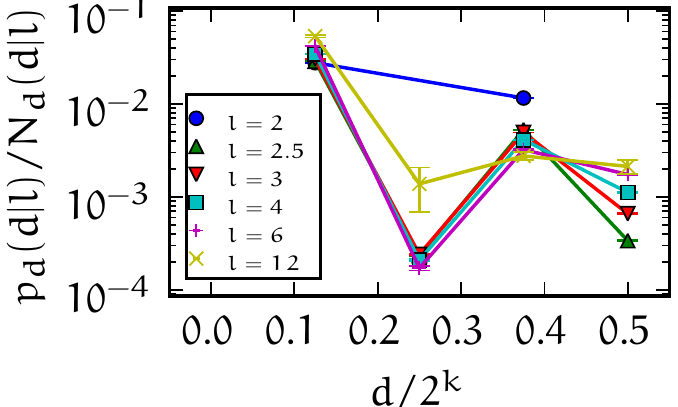}}\\
  \subfloat[$k=4$]{\includegraphics*[width=0.49\columnwidth]{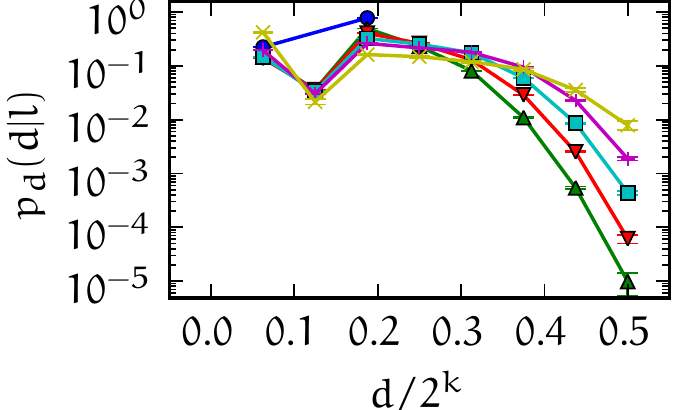}\hfill
                   \includegraphics*[width=0.49\columnwidth]{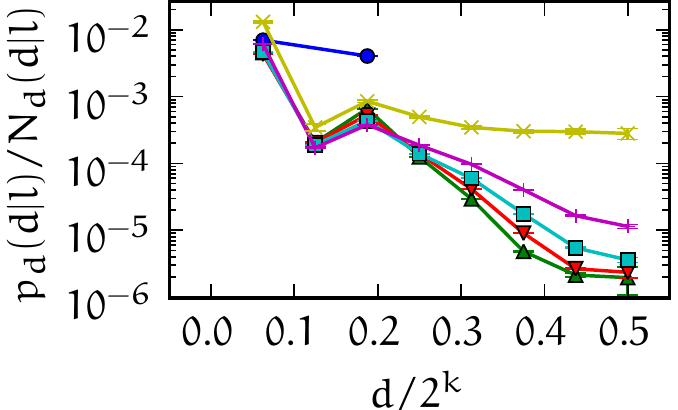}}\\
  \subfloat[$k=6$]{\includegraphics*[width=0.49\columnwidth]{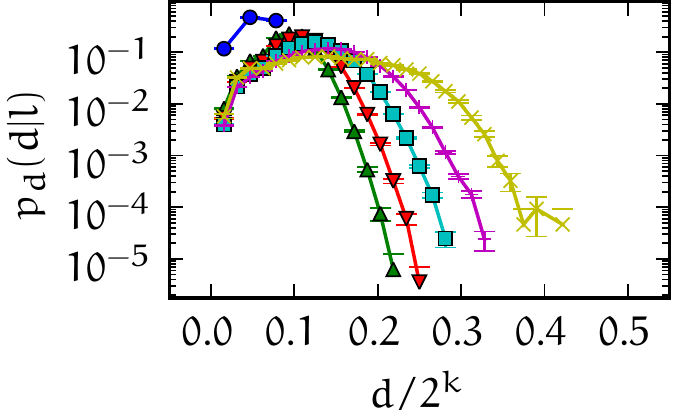}\hfill
                   \includegraphics*[width=0.49\columnwidth]{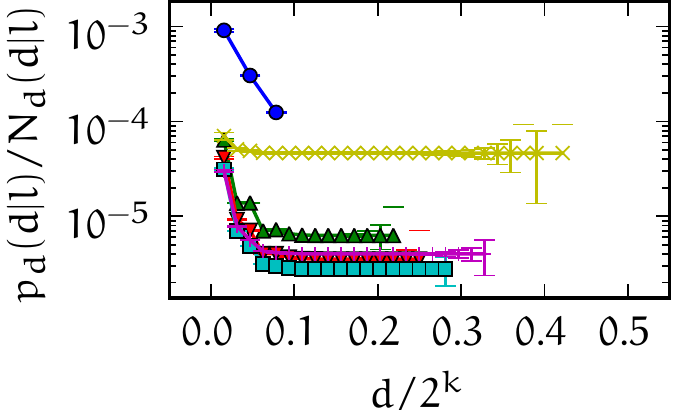}}\\
  \caption{(Color online) Distribution of functions with different values of the
    truth table homogeneity $d$, for different average flip number $l$, and
    $N=20$. \label{fig:func-classes}}
\end{figure}

Fig.~\ref{fig:func-classes} shows the probability of finding a function with a
given value of $d$. Since the number of different functions in a given class
increases rapidly with $d$ for small $d$, the maximum of this distribution is
shifted to values of $d$ larger than $1$. If this distribution is corrected by
the number $N_d$ of different functions found with the same value of $d$, an
overall decreasing function of $d$ is obtained, as shown in the graphs in the
left column of Fig.~\ref{fig:func-classes}).

The observed difference in probability due to different homogeneity
can be explained as follows.  We consider a node with $k$ inputs.  We
denote by $M=\sum_im_i \in [l/2,L-l/2]$ the total time during which
the node is in the state that it assumes less often. The sum is taken
over all intervals during which the node has this state.

If we denote the different possible (combined) states of the input nodes by
letters, we can represent the sequence of states through which the considered
node and its input states go by the following picture:
\begin{center}
 \begin{minipage}[ht]{0.4\columnwidth}
   \beginpgfgraphicnamed{fig-traj-scheme}
    \begin{tikzpicture}[rotate=120]
      \fill [color=gray] (0,0) -- (1.15cm,0) arc (360:307:1.15cm) -- cycle;
      \draw[left to reversed-right to reversed, rotate=0] (1.3cm,0) arc
      (360:307:1.3cm);
      \draw (330:1.6cm) node {$m_1$};
      \fill [color=gray, rotate=-90] (0,0) -- (1.15cm,0) arc (360:305:1.15cm) -- cycle;
      \draw[left to reversed-right to reversed, rotate=-90] (1.3cm,0) arc
      (360:305:1.3cm);
      \draw[rotate=-90] (330:1.7cm) node {$m_2$};
      \fill [color=gray, rotate=-170] (0,0) -- (1.15cm,0) arc (360:344:1.15cm) -- cycle;
      \draw[left to reversed-right to reversed, rotate=-170] (1.3cm,0) arc (360:344:1.3cm);
      \draw[rotate=-170] (355:1.7cm) node {$m_3$};
      \fill [color=gray, rotate=132] (0,0) -- (1.15cm,0) arc (360:310:1.15cm) -- cycle;
      \draw[left to reversed-right to reversed, rotate=132] (1.3cm,0) arc (360:310:1.3cm);
      \draw[rotate=132] (330:1.7cm) node {$m_i$};
      \draw [decorate,decoration={text along path,
          text={$A H F D E C H A E J K D ... F B H DJERYJ$ }}] (0.8cm,0) arc (360:0:0.8cm);
      \draw[->, thick, rotate=60] (1.3cm,0) arc (360:320:1.3cm);
      \fill [color=white] (0.75cm,0) arc (360:0:0.75cm);
    \end{tikzpicture}
    \endpgfgraphicnamed
  \end{minipage}
\end{center}
The shaded areas correspond to the output value 1. A state of the input nodes
that appears inside the shaded (clear) area, must appear again inside the shaded
(clear) area each time it is repeated. If we consider only the above scenario,
and essentially ignore that the trajectories must follow the edges of a $k$-hypercube, we
can show that functions with smaller values of $d$ should occur more often.

Our approximations rely on the fact that, for $N\to\infty$ and $l\gg 1$ (and
hence $L\to\infty)$, the shaded areas will be more numerous and will be further
apart in time and less correlated. In this limit, the input state number $i$
occurs, say, $n_i$ times.  The probability that each of the input states occurs
only in one type of area is given approximately by
\begin{equation}
   \prod_i\left[\left(\frac M L\right)^{n_i}+\left(\frac{L-M}L\right)^{n_i}\right]\, . \label{eq:M}
\end{equation}
The maximum of this function is attained at $M=l/2$ (or $M=L-l/2$, which is
excluded since we chose $M$ such that it counts the minority part), which is the
minimal possible value. The value of $d$ is bounded by $M$, but can be smaller
since the same input state can repeat. We can in fact see that the case where
the same state repeats at all $M$ times is more probable, by considering all the
possible permutations of the state sequence, for a given value of $d$,
\begin{equation}
  \left[\prod_{i\le d} {M- \displaystyle \sum_{j<i}n_j\choose n_i}\right]
  \left[\prod_{i > d} {L-M- \displaystyle \sum_{d<j<i}n_j\choose n_i}\right]
\end{equation}
and observing that it has a maximum at $d=1$, since $M\ll L$. (This
means that there are $M$ shaded areas of size 1 each.) It follows that
with increasing $l$ the weight of update functions with $d=1$ will become much
larger than that of every other update function, as is evident from
Figs.~\ref{fig:func-dist} and~\ref{fig:func-classes}. The dominance of
$d=1$ functions can already be seen for small values of $l$, although
it is less pronounced.

\FloatBarrier
\subsection{State space structure}

Finally, we investigated the state space of the constructed networks.  We
considered the system under a stochastic update scheme, since this scheme
underlies the study presented in this paper. In this case, we define an
attractor as a recurrent set of states in state space, with the property that
there are no transitions that escape this set (i.e. a strongly connected
component in the state space graph that has no outgoing connections). The number
of states in this set is called the size of the attractor.

\begin{figure}[htb]
  \includegraphics*[width=\columnwidth]{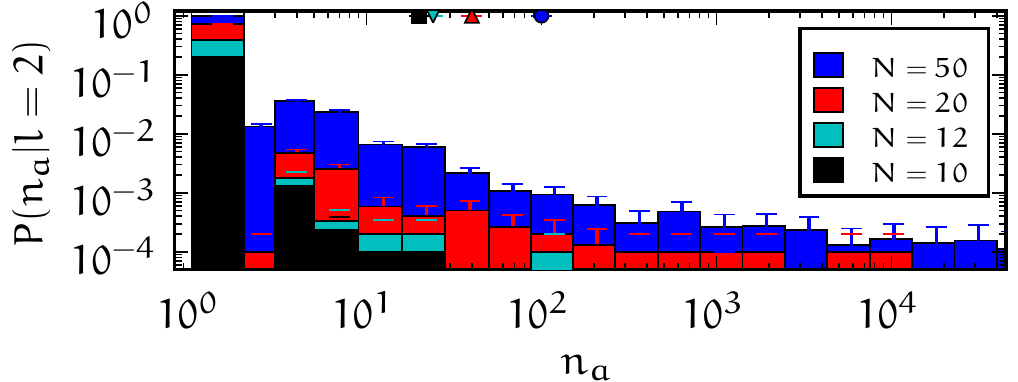}
  \includegraphics*[width=\columnwidth]{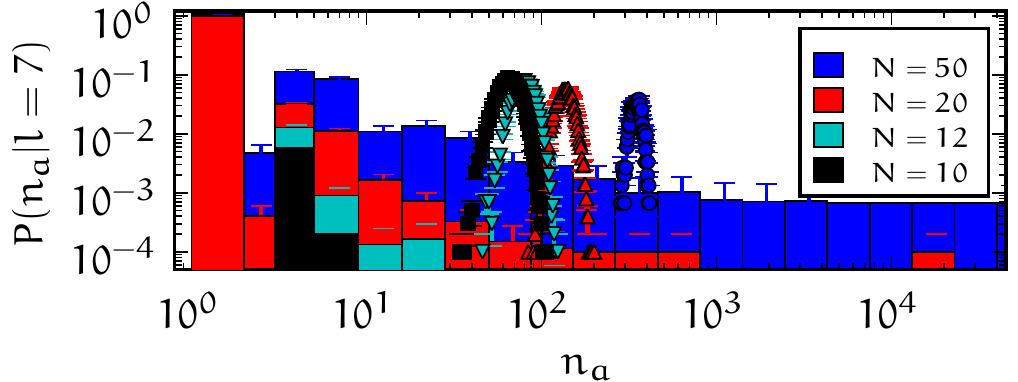}
\caption{(Color online) Probability of attractor sizes $n_a$, for $l=2$ and
  $l=7$. Attractors corresponding to the given trajectories are plotted
  separately with symbols. For each value of $N$ and $l$, $10^4$ different
  networks were analysed. In the case of attractor sampling, $100$ different
  random initial conditions per network were used. \label{fig:attractorsizes}}
\end{figure}

We evaluated the probabilities of attractors of a given size on the ensemble of
minimal networks, and their average basin size. For small networks (up to
$N=12$) the attractor size probability was obtained by exact enumeration of the
state space. For larger $N$, the state space was sampled, taking care that the
same attractor was not counted twice. This method, however, leads to a bias,
since attractors with smaller basins are less likely to be counted, and the
extent of this bias depends on the size of the network. Nevertheless, this bias
is not relevant for our point of interest, which is on the occurrence of various
attractors, but not on their precise statistics.  Fig.~\ref{fig:attractorsizes}
shows that there exist almost always fixed points, and that there are often
attractors which are much larger than the imposed reliable trajectory (we
considered attractors of up to $n_a=10^5$ states). Note that the probabilities
in Fig.~\ref{fig:attractorsizes} do not sum to $1$, since a given network may
have many attractors of different sizes.

\begin{figure}[htb]
  \includegraphics*[width=\columnwidth]{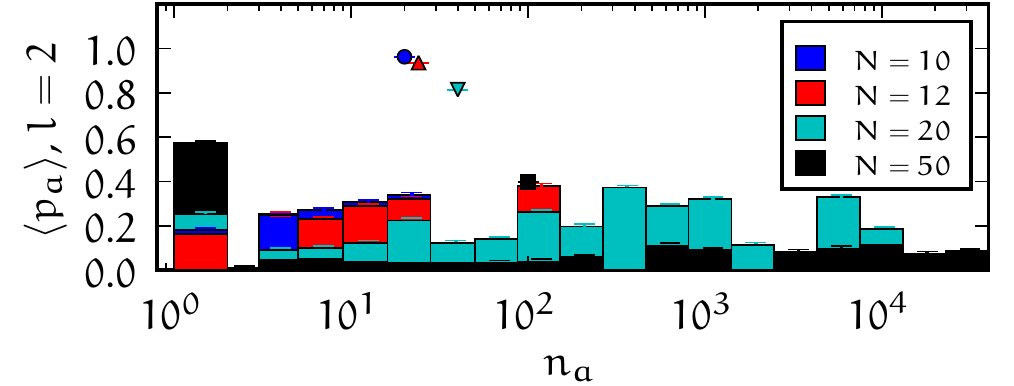}\\
  \includegraphics*[width=\columnwidth]{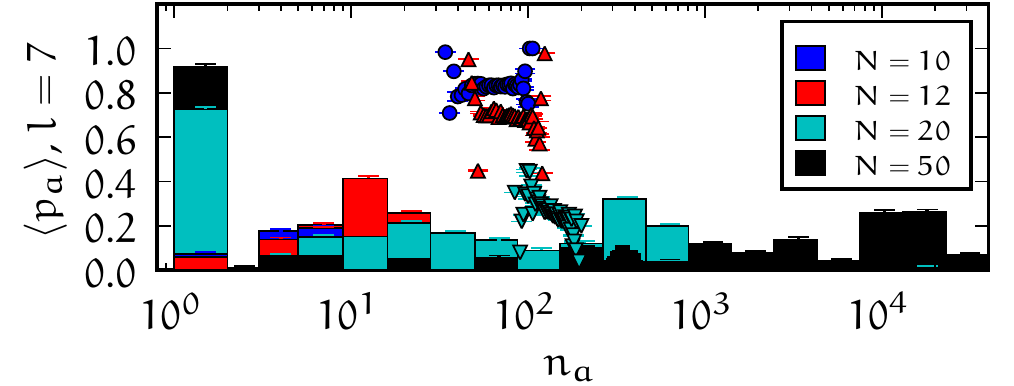}
\caption{(Color online) Average attractor probabilities (basin size),
  $\left<p_a\right>$, for $l=2$ and $l=7$. Attractors corresponding to the given
  trajectories are plotted separately with symbols. For each value of $N$ and
  $l$, $10^4$ different networks were analysed, and $100$ different random
  initial conditions per network were used.\label{fig:basinsizes}}
\end{figure}

The basin of attraction was measured as the probability of reaching an
attractor, starting from a random configuration. Fig.~\ref{fig:basinsizes} shows
that the omnipresent fixed point has a large basin of attraction. Larger
attractors occur with smaller probabilities.  The weight of the fixed point,
compared to the weight of the imposed reliable trajectory, increases with
increasing $N$.  This can be explained by the entries in the truth table which
are not uniquely determined by the reliable trajectory: While number of entries
fixed throughout the trajectory grow linearly with $l$, the number of remaining
entries (as well as their contribution to the state space) grow
exponentially. In this increasingly large region of the state space, the
functions behave as constant functions.

Attractors which are larger than the given trajectories are due to a
portion of network begin frozen in the value they have at the fixed
point, while other nodes remain frustrated, and their states change
stochastically, visiting a larger portion of the state space, without
entering the fixed point or the reliable trajectory. 

For comparison, we briefly looked at the attractor sizes obtained
using a synchronous updating scheme. Not surprisingly, the attractors
become much shorter in this case, with attractors larger than the
given trajectory having only a small probability (not shown).

\FloatBarrier
\section{Conclusion}

We have constructed minimal Boolean networks which follow a given
reliable trajectory in state space. The trajectories considered have
the necessary feature that only one node can change its value at any
moment in time, which guarantees that the sequence of states is
independent of the order in which nodes are updated. Otherwise the nodes
change their states at randomly assigned times in the given
trajectory, thus constituting a null model for reliable dynamics. The
minimality condition imposed on the networks was that it contains the
smallest possible set of inputs for each node that allows for the
given trajectory. Additionally, the truth table entries that are not
fixed by the trajectory were set to the majority value imposed by the
trajectory.  We then investigated the topology, the update functions,
and the state space of those networks.

The network structure, as manifest in the degree distribution, does
not deviate significantly from a random topology. However, the network
exhibits larger clustering than a random network, and exhibits a
characteristic motif profile, which resembles both real networks of
gene regulation and the pattern of dynamically reliable motifs found
in~\cite{klemm_topology_2005}. The existence of clustering and motifs
was explained by considering the ``excess'' inputs that are required
to avoid contradictions in the truth table, and how they must
be correlated among each other.

The update functions of the nodes show a characteristic distribution,
where only a subset of the possible functions occur, and these are
divided into distinct classes, which occur with different
probabilities. The main factor discerning the different classes is
their homogeneity, characterized by the number of entries of the
minority bit in the truth table. Function with homogeneity 1 occur
with increased probability, and become the dominant functions in the
limit of large trajectories, $l\to\infty$, for fixed $k$. Functions
with more minority entries occur with a smaller probability, and this
probability decreases as the number of minority entries increase. We
presented an analytical justification for this finding, considering how
the local trajectory of the input states of a given function must behave, in
the limit $l\gg 1$.

Finally we investigated the state space of the constructed networks,
considering the possible attractors it can have, in addition to the
given reliable trajectory. To this aim, we used a stochastic update
scheme. We observed that the network almost always exhibits a fix
point of the dynamics, and often attractors which can be much larger
than the given trajectory. The basin size of the fixed point is very
large, and dominates the basin size of the given trajectory in the
limit of large system size. This is a consequence of the minimality
condition imposed on the network: The region of state space dictated
by the imposed trajectory increases only linearly with system size,
while the entire state space grows exponentially. Outside the state
space region fixed by the reliable trajectory, the constructed
functions behave as constant functions, which drive the system nearer
to the frozen phase.

In this work, we have used a null model for reliable trajectories,
where the nodes change their values at random times. Real gene
regulatory networks deviate significantly from this, since they must
agree with the cell cycle or the pathway taken during embryonic
development.  Certain proteins need to be always present in the cell,
while others are produced only under specific conditions. The degree
distribution and the update functions must reflect this behavior.
However, some of the features found for the null model presented in
this work, should also be present in more realistic systems. The
existence of clustering, and the motif profile found, for instance, do
not depend strongly on the specific temporal patterns of the nodes,
but are imposed by the reliability condition. Similarly, the dominance
of strongly canalyzing functions is a consequence of the reliability
condition and should be relatively robust to the introduction of temporal
correlations. Nevertheless, biochemistry  makes some canalyzing
functions more likely than others.

An important feature of biological networks that is not reflected in
the null model presented in this paper, is the robustness with respect
to perturbations of a node. Such a robustness can only be obtained
when more than the minimum possible number of inputs is assigned to a
node. Indeed, it has been shown in~\cite{gershenson_role_2006} that more
redundancy allows for more robustness. Similarly, requiring that the
reliable trajectory has the largest basin of attraction or that other
attractors of the system are also reliable trajectories, may increase
the number of links in the network.

Finally, the requirement that trajectories are fully reliable is an idealization
which goes beyond what is necessary for gene regulatory networks. Real networks
have checkpoint states, but between these states, the precise sequence of events
is not always important. On the other hand, full reliability may be necessary
for certain subsystems of the gene network, where a strict sequence of local
states is required. The minimal reliable networks discussed in this paper should
be compared more realistically to such reliable modules.


\section{Acknowledgements}

We thank Ron Milo for providing the signal-transduction network data. We
acknowledge the support of this work by the Humboldt Foundation and by the DFG
under contract number Dr300/5-1.

\bibliography{bib}

\end{document}